\documentclass[12pt]{article}
\usepackage{epsf}

\usepackage{amssymb}
\usepackage{amsmath}
\usepackage{graphicx}
\usepackage{latexsym}

\begin{document}

\def\ds{\displaystyle}
\def\Slash#1{{\ooalign{\hfil$#1$\hfil\crcr\hfil$/$\hfil}}}

\newcommand{\rem}[1]{{\bf #1}}

\renewcommand{\theequation}{\thesection.\arabic{equation}}

\renewcommand{\thefootnote}{\fnsymbol{footnote}}
\setcounter{footnote}{0}
\begin{titlepage}

\def\thefootnote{\fnsymbol{footnote}}

\begin{center}

\hfill UT-HET 010 \\
\hfill TU-820\\
\hfill June, 2008\\

\vskip .75in

{\Large \bf 
  Testing the Littlest Higgs Model with T-parity\\
at the Large Hadron Collider
}
\vskip .75in

{\large
$^{(a)}$Shigeki Matsumoto, $^{(b)}$Takeo Moroi, and $^{(c)}$Kazuhiro Tobe
}

\vskip 0.25in

{\em $^{(a)}$Department of Physics, University of Toyama, 
    Toyama 930-8555, Japan}

\vskip 0.2in

{\em $^{(b)}$Department of Physics, Tohoku University,
Sendai 980-8578, Japan}

\vskip 0.2in

{\em $^{(c)}$Department of Physics, Nagoya University,
Nagoya 464-8602, Japan}

\end{center}
\vskip .5in

\begin{abstract}

  In the framework of the littlest Higgs model with T-parity (LHT), we
  study the production processes of T-even ($T_+$) and T-odd ($T_-$)
  partners of the top quark at the Large Hadron Collider (LHC).  We show
  that the signal events can be distinguished from the standard-model
  backgrounds, and that information about mass and mixing parameters
  of the top partners can be measured with relatively good accuracies.
  With the measurements of these parameters, we show that a
  non-trivial test of the LHT can be performed.  We also discuss a
  possibility to reconstruct the thermal relic density of the lightest T-odd
  particle $A_H$ using the LHC results, and show that the scenario
  where $A_H$ becomes dark matter may be checked.

 \end{abstract}

\end{titlepage}

\renewcommand{\thepage}{\arabic{page}}
\setcounter{page}{1}
\renewcommand{\thefootnote}{\#\arabic{footnote}}
\setcounter{footnote}{0}

\section{Introduction}
\label{sec:intro}
\setcounter{equation}{0}

The hierarchy problem in the standard model (SM) is expected to give a
clue to explore physics beyond the SM. This problem is essentially
related to quadratically divergent corrections to the Higgs boson
mass, and it strongly suggests the existence of new physics at the TeV
scale. At the new physics scale, the problem is expected to be
resolved due to the appearance of a new symmetry which controls the
Higgs boson mass. With this philosophy, a lot of scenarios have been
proposed so far. The most famous example is the supersymmety (SUSY),
by which quadratically divergent corrections to the Higgs boson mass are
completely cancelled. Another example is the Gauge-Higgs unification,
by which the gauge invariance in higher dimensional space-time
protects the Higgs potential from any ultraviolet (UV) divergent
corrections.

In this article, we consider the third possibility, so-called the
little Higgs (LH) scenario \cite{ref:LH}, in which the Higgs boson
mass is controlled by a global symmetry. In this scenario, the Higgs
boson is regarded as a pseudo Nambu-Goldstone boson arising from the
spontaneous breaking of a symmetry. Due to the symmetry
imposed, new particles such as heavy gauge bosons and top-partners are
necessarily introduced, and main quadratically divergent corrections to
the Higgs boson mass vanish at one-loop level due to contributions of
these particles.  Unlike the SUSY scenario, the cancellation of the
quadratic divergence is achieved only at one-loop level, thus
the LH model needs a UV completion at some higher scale.
However, due to the cancellation at one-loop level, the fine-tuning of
the Higgs boson mass is avoided even if the cutoff scale of the LH
model is around 10 TeV. As a result, the LH model solves the little
hierarchy problem \cite{ref:LittleHierarchy}.

Unfortunately, the original LH model is severely constrained by
electroweak precision measurements due to direct couplings among a new
heavy gauge boson and SM particles \cite{ref:Difficulty}. In order to
resolve the problem, the implementation of the $Z_2$ symmetry called
T-parity to the model has been proposed
\cite{Cheng:2003ju,Cheng:2004yc,Low:2004xc}.  Under the parity, almost
all new particles are T-odd, while the SM particles are
T-even\footnote{One important exception is the top-partner
  $T_+$, which is a T-even new particle as shown in the next
  section.}. Thanks to the symmetry, dangerous interactions stated
above are prohibited \cite{Hubisz:2005tx}. Furthermore, the lightest
T-odd particle (LTP) becomes stable, which is electrically and color
neutral, and has a mass of ${\cal O}$(100) GeV in many little Higgs
models with T-parity \cite{Cheng:2003ju}. Therefore, these models
provide a good candidate for dark matter \cite{ref:LHdarkmatter}\footnote{For
  UV completion of T-parity models, see \cite{ref:UVforLH}.}.
  
In this article, we study signatures of the littlest Higgs model with
T-parity (LHT) \cite{Cheng:2004yc,Low:2004xc} at the Large Hadron
Collider (LHC), which is expected to explore various new-physics
models \cite{AtlasTdr,CmsTdr}. The LHT is the simplest model
realizing the LH scenario with the T-parity, and considered to be an
attractive reference model. Since the LHC is a hadron collider, new
colored particles have an important role to explore physics beyond the
SM. As shown in the next section, top-partners are necessarily
introduced in the LH models, which are responsible for the
cancellation of quadratically divergent corrections to the Higgs boson
mass from top loop diagrams. Furthermore, masses of these partners are
expected to be less than $\sim$ 1 TeV, and the partners will be
copiously produced at the LHC \cite{ref:LHatLHC}. Therefore, we
consider the productions of the top partners at the LHC with a
realistic simulation study, and show that these signatures are clearly
distinguishable from SM backgrounds. Furthermore, we find that it is
also possible to test the LHT by investigating a non-trivial relation
among the signatures.  We also consider how accurately model
parameters of the LHT are determined, and discuss its implication to
the property of the LTP dark matter such as how precisely the relic
abundance of the dark matter is estimated with the LHC data.

This paper is organized as follows. In the next section, we briefly
review the littlest Higgs model with T-parity paying particular
attention to the gauge-Higgs and top sectors of the model. We also
present representative points used in our simulation study. Signatures
of the LHT at the LHC are shown in Sec.\ \ref{sec:tplus}, especially
focusing on the pair production of T-even top partner, the single
production of T-even top partner, and the pair production of T-odd
partner. The test of the LHT is discussed in Sec.\ \ref{sec:test},
where we investigate a non-trivial relation among the signatures
obtained in the previous section. We also discuss the implication of
the result to the LTP dark matter phenomenology. Sec.\
\ref{sec:summary} is devoted to summary.

\section{Model}
\label{sec:model}

In this section, we briefly review the littlest Higgs model with
T-parity focusing on gauge-Higgs and top sectors of the model. (For
general reviews of little Higgs models and their phenomenological
aspects, see \cite{ref:LHreview,ref:PhenomenologyLH}.) We also
present a few representative points used in our simulation study at
the end of this section.

\subsection{Gauge-Higgs sector}

The littlest Higgs model with T-parity is based on a non-linear sigma
model describing an SU(5)/SO(5) symmetry breaking. The non-linear
sigma field $\Sigma$ is given as
\begin{eqnarray}
 \Sigma = e^{2i\Pi/f}\Sigma_0,
\end{eqnarray}
where $f \sim {\cal O}(1)$ TeV is the vacuum expectation value of the
breaking. The Nambu-Goldstone (NG) boson matrix $\Pi$ and the
direction of the breaking $\Sigma_0$ are
\begin{eqnarray}
 \Pi
 =
 \begin{pmatrix}
                   0 & H  /\sqrt{2} & \Phi         \\
  H^\dagger/\sqrt{2} &            0 & H^T/\sqrt{2} \\
        \Phi^\dagger & H^*/\sqrt{2} & 0            \\
 \end{pmatrix},
 \qquad
 \Sigma_0
 =
 \begin{pmatrix}
        0 & 0 & {\bf 1} \\
        0 & 1 &       0 \\
  {\bf 1} & 0 &       0 \\
 \end{pmatrix}.
 \label{pNG matrix}
\end{eqnarray}
Here, we omit the would-be NG fields in the $\Pi$ matrix. An
[SU(2)$\times$U(1)]$^2$ subgroup in the SU(5) global symmetry is
gauged, which is broken down to the diagonal subgroup identified with
the SM gauge group SU(2)$_L\times$U(1)$_Y$. Due to the presence of the
gauge interactions and Yukawa interactions introduced in the next
subsection, the SU(5) global symmetry is not exact, and particles in
the $\Pi$ matrix become pseudo NG bosons. Fourteen (= 24 $-$ 10) NG
bosons are decomposed into representations ${\bf 1}_0 \oplus {\bf 3}_0
\oplus {\bf 2}_{\pm 1/2} \oplus {\bf 3}_{\pm 1}$ under the electroweak
gauge group. The first two representations are real, and become
longitudinal components of heavy gauge bosons when the
[SU(2)$\times$U(1)]$^2$ is broken down to the SM gauge group. The
other scalars ${\bf 2}_{\pm 1/2}$ and ${\bf 3}_{\pm 1}$ are a complex
doublet identified with the SM Higgs field ($H$ in Eq.\ (\ref{pNG
  matrix})) and a complex triplet Higgs field ($\Phi$ in Eq.\ (\ref{pNG
  matrix})), respectively.

The kinetic term of the $\Sigma$ field is given as
\begin{eqnarray}
  {\cal L}_{\Sigma}
  =
  \frac{f^2}{8}{\rm Tr}
  \left|
    \partial_\mu \Sigma
    -
    i\sqrt{2}
    \left\{
      g  ({\bf W} \Sigma + \Sigma {\bf W}^T)
      +
      g' ({\bf B} \Sigma + \Sigma {\bf B}^T)
    \right\}
  \right|^2,
  \label{Kinetic}
\end{eqnarray}
where ${\bf W} = W^a_j Q_j^a$ (${\bf B} = B_j Y_j$) is the ${\rm SU(2)}_j$
(${\rm U(1)}_j$) gauge field and $g$ ($g'$) is the ${\rm SU(2)}_L$ (${\rm U(1)}_Y$)
gauge coupling constant. With the Pauli matrix $\sigma^a$, the
generator $Q_j$ and the hyper-charge $Y_j$ are given as
\begin{eqnarray}
  &&
  Q_1^a = +\frac{1}{2}
  \begin{pmatrix}
    \sigma^a & 0 & 0 \\
    0 & 0 & 0 \\
    0 & 0 & 0
  \end{pmatrix},
  \qquad
  Y_1 = {\rm diag}(3,3,-2,-2,-2)/10,
  \\
  &&
  Q_2^a = -\frac{1}{2}
  \begin{pmatrix}
    0 & 0 & 0 \\
    0 & 0 & 0 \\
    0 & 0 & \sigma^{a*}
  \end{pmatrix},
  \qquad
  Y_2 = {\rm diag}(2,2,2,-3,-3)/10.
\end{eqnarray}
It turns out that the Lagrangian in Eq.\ (\ref{Kinetic}) is invariant
under the T-parity,
\begin{eqnarray}
  \Pi \leftrightarrow -\Omega \Pi \Omega,
  \qquad
  W^a_1 \leftrightarrow W^a_2,
  \qquad
  B_1 \leftrightarrow B_2,
\end{eqnarray}
where $\Omega = {\rm diag}(1,1,-1,1,1)$.

This model contains four kinds of gauge fields. The linear
combinations $W^a = (W^a_1 + W^a_2)/\sqrt{2}$ and $B = (B_1 +
B_2)/\sqrt{2}$ correspond to the SM gauge bosons for the SU(2)$_L$ and
U(1)$_Y$ symmetries. The other linear combinations $W^a_H = (W^a_1 -
W^a_2)/\sqrt{2}$ and $B_H = (B_1 - B_2)/\sqrt{2}$ are additional gauge
bosons, which acquire masses of ${\cal O}(f)$ through the SU(5)/SO(5)
symmetry breaking. After the electroweak symmetry breaking with
$\langle H \rangle = (0, v/\sqrt{2})^T$, the neutral components of
$W^a_H$ and $B_H$ are mixed with each other and form mass eigenstates
$A_H$ and $Z_H$,
\begin{eqnarray}
  \begin{pmatrix}
    Z_H \\ A_H
  \end{pmatrix}
  =
  \begin{pmatrix}
    \cos \theta_H & -\sin \theta_H \\
    \sin \theta_H &  \cos \theta_H     
  \end{pmatrix}    
  \begin{pmatrix}
    W_H^3 \\ B_H 
  \end{pmatrix}.
\end{eqnarray}
The mixing angle $\theta_H$ is given as
\begin{eqnarray}
  \tan \theta_H
  =
  - \frac{2m_{12}}
  {m_{11} - m_{22} + \sqrt{(m_{11} - m_{22})^2 + 4m_{12}^2}}
  \sim
  - 0.15\frac{v^2}{f^2},
\end{eqnarray}
where $m_{11} = g^2 f^2 (c_f^2 + 7)/8$, $m_{12} = g g^{\prime } f^2 (1
- c_f^2)/8$, $m_{22} = g^{\prime 2} f^2 (5c_f^2 + 3)/40$, and $c_f =
\cos (\sqrt{2}v/f)$. Since the mixing angle is considerably
suppressed, $A_H$ is dominantly composed of $B_H$. Masses of gauge
bosons are
\begin{eqnarray}
  m_W^2 &=& \frac{g^2}{4} f^2 (1 - c_f)
  \simeq \frac{g^2}{4} v^2,
  \\
  m_Z^2 &=& \frac{g^2 + g^{\prime 2}}{4} f^2 (1 - c_f)
  \simeq \frac{g^2 + g^{\prime 2}}{4} v^2,
  \\
  m_{W_H}^2 &=& \frac{g^2}{4} f^2 (c_f + 3)
  \simeq g^2 f^2,
  \\
  m_{Z_H}^2
  &=&
  \frac{1}{2}
  \left(m_{11} + m_{22} + \sqrt{(m_{11} - m_{22})^2 + 4m_{12}^2}\right)
  \simeq g^2 f^2,
  \\
  m_{A_H}^2
  &=&
  \frac{1}{2}
  \left(m_{11} + m_{22} - \sqrt{(m_{11} - m_{22})^2 + 4m_{12}^2}\right)
  \simeq 0.2 g^{\prime 2} f^2.
  \label{m(A_H)}
\end{eqnarray}
As expected from the definitions of $A_H$, $Z_H$, and $W_H$, the new
heavy gauge bosons behave as T-odd particles, while SM gauge bosons
are T-even.

A potential term for $H$ and $\Phi$ fields is radiatively generated as
\cite{ref:LH,ref:LHdarkmatter}
\begin{eqnarray}
 V(H, \Phi)
 =
 \lambda f^2{\rm Tr}\left[\Phi^\dagger\Phi\right]
 -
 \mu^2H^\dagger H
 +
 \frac{\lambda}{4}\left(H^\dagger H\right)^2
 +
 \cdots.
 \label{Potential}
\end{eqnarray}
Main contributions to $\mu^2$ come from logarithmic divergent
corrections at 1-loop level and quadratically divergent corrections at
2-loop level. As a result, $\mu^2$ is expected to be smaller than
$f^2$. The triplet Higgs mass term, on the other hand, receives
quadratically divergent corrections at 1-loop level, and therefore is
proportional to $f^2$. The quartic coupling $\lambda$ is determined by
the 1-loop effective potential from gauge and top sectors. Since both
$\mu$ and $\lambda$ depend on parameters at the cutoff scale $\Lambda
\simeq 4\pi f$, we treat them as free parameters in this paper. The
mass of the triplet Higgs boson $\Phi$ is given by $m_\Phi^2 = \lambda
f^2 = 2m_h^2f^2/v^2$, where $m_h$ is the mass of the SM Higgs
boson. The triplet Higgs boson is T-odd, while the SM Higgs is T-even.

Gauge-Higgs sector of the LHT is composed of the kinetic term of
$\Sigma$ field in Eq.\ (\ref{Kinetic}) and the potential term in Eq.\
(\ref{Potential}) in addition to appropriate kinetic terms of gauge
fields $W^a_j$, $B_j$ and gluon $G$. It can be seen that 
the heavy photon $A_H$ is considerably
lighter than other T-odd particles. Since the stability of $A_H$ is
guaranteed by the conservation of T-parity, it becomes a good
candidate for dark matter.

\subsection{Top sector}

To implement T-parity, two SU(2) doublets $q^{(1)}$ and $q^{(2)}$ and
one singlet $u_R$ are introduced for each SM fermion. Furthermore, two
vector-like singlets $U^{(1)}$ and $U^{(2)}$ are also introduced in
the top sector in order to cancel large radiative corrections to the
Higgs mass term. The quantum numbers of the particles in the top
sector under the [SU(2)$\times$ U(1)]$^2$ gauge symmetry are shown in
Table \ref{table:charges}.  All particles are triplets under the SM
SU(3)$_c$ (color) symmetry.

\begin{table}[t]
  \center{
    \begin{tabular}{|c|c||c|c|}
      \hline
      $q^{(1)}$ & $({\bf 2}, 1/30; {\bf 1}, 2/15)$ &
      $q^{(2)}$ & $({\bf 1}, 2/15; {\bf 2}, 1/30)$ \\
      \hline
      $U^{(1)}_L$ & $({\bf 1}, 8/15; {\bf 1}, 2/15)$ &
      $U^{(2)}_L$ & $({\bf 1}, 2/15; {\bf 1}, 8/15)$ \\
      \hline
      $U^{(1)}_R$ & $({\bf 1}, 8/15; {\bf 1}, 2/15)$ &
      $U^{(2)}_R$ & $({\bf 1}, 2/15; {\bf 1}, 8/15)$ \\
      \hline
      $u_R$ & $({\bf 1}, 1/3 ; {\bf 1}, 1/3 )$ &
      &                                  \\
   \hline
  \end{tabular}
  }
  \caption{\small Quantum number for $[SU(2)\times U(1)]^2$ 
    for particles in the top sector.}
  \label{table:charges}
\end{table}

With these particles, Yukawa interactions which are invariant under
gauge symmetries and T-parity turn out to be
\begin{eqnarray}
 {\cal L}_t
 =
 \frac{\lambda_1 f}{2\sqrt{2}} \epsilon_{ijk} \epsilon_{xy}
 \left[
  \left(\bar{\cal Q}^{(2)} \Sigma_0\right)_i
  \tilde{\Sigma}_{j x} \tilde{\Sigma}_{k y}
  -
  \bar{\cal Q}^{(1)}_i \Sigma_{j x} \Sigma_{k y}
 \right] u_R
 - \lambda_2 f \sum_{n = 1}^2 \bar{U}^{(n)}_L U^{(n)}_R + {\rm h.c.},
 \label{top yukawa}
\end{eqnarray}
where ${\cal Q}^{(n)} = (q^{(n)}, U^{(n)}_L, 0)^T$, $q^{(n)} =
-\sigma^2 (u^{(n)}_L, b^{(n)}_L)^T$, and $\tilde{\Sigma} =
\Sigma_0\Omega\Sigma^\dagger\Omega\Sigma_0$.  The indices $i,j,k$ run
from 1 to 3, while $x,y = 4,5$. The coupling constant $\lambda_1$ is
introduced to generate the top Yukawa coupling and $\lambda_2 f$ gives the
vector-like mass of the singlet $U^{(n)}$. Under T-parity, $q^{(n)}$
and $U^{(n)}$ transform as $q^{(1)} \leftrightarrow - q^{(2)}$ and
$U^{(1)} \leftrightarrow - U^{(2)}$, thus T-parity eigenstates are
given as
\begin{eqnarray}
 q^{(\pm)}
 =
 \frac{1}{\sqrt{2}} \left( q^{(1)} \mp q^{(2)} \right),
 \qquad
 U^{(\pm)}_{L(R)}
 =
 \frac{1}{\sqrt{2}} \left( U^{(1)}_{L(R)} \mp U^{(2)}_{L(R)} \right).
\end{eqnarray}

In terms of the eigenstates, mass terms in Eq.\ (\ref{top yukawa}) are
written as
\begin{eqnarray}
  {\cal L}_{\rm mass}
  =
  - \lambda_1
  \left[ f\bar{U}^{(+)}_L + v \bar{u}^{(+)}_L \right] u_R
  - \lambda_2 f
  \left( \bar{U}^{(+)}_L U^{(+)}_R + \bar{U}^{(-)}_L U^{(-)}_R \right)
  + {\rm h.c.}
\end{eqnarray}
T-even states $u_+$ and $U_+$ form the following mass eigenstates
\begin{eqnarray}
 \begin{pmatrix}
  t_L \\ T_{+L}
 \end{pmatrix}
 =
 \begin{pmatrix}
  \cos\beta & -\sin\beta \\
  \sin\beta &  \cos\beta
 \end{pmatrix}
 \begin{pmatrix}
  u^{(+)}_L \\ U^{(+)}_L
 \end{pmatrix},
 ~~~
 \begin{pmatrix}
  t_R \\ T_{+R}
 \end{pmatrix}
 =
 \begin{pmatrix}
  \cos\alpha & -\sin\alpha \\
  \sin\alpha &  \cos\alpha
 \end{pmatrix}
 \begin{pmatrix}
  u^{(+)}_R \\ U^{(+)}_R
 \end{pmatrix}.
\end{eqnarray}
Mixing angles $\alpha$, $\beta$ and mass eigenvalues $m_t$, $m_{T_+}$
are given as
\begin{eqnarray}
 \tan \alpha
 &=&
 \frac{2 B_t C_t}{\Delta_t - (A_t^2 + B_t^2 - C_t^2)}
 \simeq
 \lambda_1/\lambda_2,
 \nonumber \\
 \tan \beta
 &=&
 \frac{2 A_t B_t}{\Delta_t - (A_t^2 - B_t^2 - C_t^2)}
 \simeq
 \frac{\lambda_1^2}{\lambda_1^2 + \lambda_2^2} \frac{v}{f},
 \nonumber \\
 m_t
 &=&
 \frac{1}{\sqrt{2}} \sqrt{A_t^2 + B_t^2 + C_t^2 - \Delta_t}
 \simeq
 \frac{\lambda_1 \lambda_2}{\sqrt{\lambda_1^2 + \lambda_2^2}} v,
 \nonumber \\
 m_{T_+}
 &=&
 \frac{1}{\sqrt{2}} \sqrt{A_t^2 + B_t^2 + C_t^2 + \Delta_t}
 \simeq
 \sqrt{\lambda_1^2 + \lambda_2^2} f,
\end{eqnarray}
where $A_t = s_f \lambda_1 f/\sqrt{2}$, $B_t = (1 + c_f)\lambda_1
f/2$, $C_t = \lambda _2 f$, and $\Delta_t = ((A_t^2 + B_t^2 + C_t^2)^2
- 4 A_t^2 C_t^2)^{1/2}$ with $s_f$ being $s_f =
\sin(\sqrt{2}v/f)$. The $t$ quark is identified with the SM top quark,
and $T_+$ is its T-even heavy partner. On the other hand, the T-odd
fermions $U_{L-}$ and $U_{R-}$ form a Dirac fermion, $T_-$, whose mass
is given by $m_{T_-} = \lambda_2f$. The remaining T-odd quark $q_-$
acquires mass by introducing an additional SO(5) multiplet
transforming nonlinearly under the SU(5) symmetry. Therefore, the mass
term of the quark does not depend on $\lambda_1$ and $\lambda_2$. In
this paper, we assume that the $q_-$ quark is heavy enough compared to
other top partners, and that it is irrelevant for the direct production at the LHC
experiment. (For the phenomenology of the $q_-$ quark, see
\cite{Belyaev:2006jh}.)  Finally, it is worth notifying that the T-odd
partner of top quark ($T_-$) does not participate in the cancellation
of quadratically divergent corrections to the Higgs mass term. The
cancellation is achieved by only loop diagrams involving $t$ and $T_+$
quarks.

\subsection{Representative points}

In this paper, we focus on $T_{\pm}$ productions at the LHC. For this purpose,
we need to choose representative points to perform a numerical
simulation. In order to find attractive points, we consider those
consistent with electroweak precision measurements and the WMAP
experiment for dark matter relics\footnote{We consider the WMAP
  constraint only for choosing a representative point. In fact, the
  model does not have to satisfy the constraint, because, for
  instance, dark matter may be composed of other particles such as the
  axion.}.

We consider a $\chi^2$-function to choose representative points;
\begin{eqnarray}
 \chi^2
 =
 \sum_i
 \frac{\left({\cal O}_{\rm obs}^{(i)} - {\cal O}_{\rm th}^{(i)}\right)^2}
      {\left(\Delta {\cal O}_{\rm obs}^{(i)}\right)^2},
 \label{chi2}
\end{eqnarray}
where ${\cal O}_{\rm obs}^{(i)}$, ${\cal O}_{\rm th}^{(i)}$, and
$\Delta {\cal O}_{\rm obs}^{(i)}$ are experimental result, theoretical
prediction, and the error of the observation for observable ${\cal
  O}$. We consider following eight observables; $W$ boson mass ($m_W
=$ 80.412$\pm$0.042 GeV), weak mixing angle ($\sin^2\theta^{\rm
  lept}_{\rm eff} =$ 0.23153$\pm$0.00016), leptonic width of the $Z$
boson ($\Gamma_l =$ 83.985$\pm$0.086 MeV) \cite{LEPSLC}, fine
structure constant at the $Z$ pole ($\alpha^{-1}(m_Z) =$
128.950$\pm$0.048), top quark mass ($m_t =$ 172.7$\pm$2.9 GeV)
\cite{Arguin:2005cc}, $Z$ boson mass ($m_Z =$ 91.1876$\pm$0.0021 GeV),
Fermi constant ($G_F =$ (1.16637$\pm$0.00001)$\times$10$^{-5}$
GeV$^{-2}$) \cite{Yao:2006px}, and relic abundance of dark matter
($\Omega_{\rm DM} h^2 =$ 0.119$\pm$0.009) \cite{de Austri:2006pe}. 
On the other hand,
theoretical predictions of these observables depend on seven model
parameters; $f$, $\lambda_2$, $m_h$, $\alpha^{-1}(m_Z)$, $G_F$, $m_Z$,
and $m_t$. (For the detailed expressions of the theoretical
predictions, see \cite{Hubisz:2005tx,ref:LHdarkmatter}). In order to
obtain the constraint on $f$ vs.\ $\lambda_2$ plane, we minimize the
$\chi^2$ function in Eq.\ (\ref{chi2}) with respect to parameters
$m_h$, $\alpha^{-1}(m_Z)$, $G_F$, $m_Z$, and $m_t$. In other words, we
integrate out these parameters from the probability function $P \equiv
e^{-\chi^2/2}$.

% Fig.1
\begin{figure}[t]
 \begin{center}
  \scalebox{0.8}{\includegraphics*{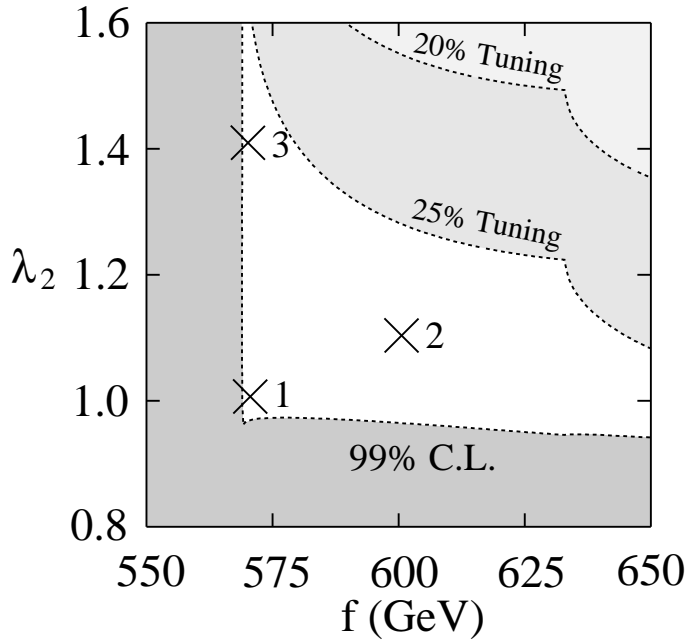}}
  \caption{\small Constraints to the littlest Higgs model with
    T-parity on $f$ vs.\ $\lambda_2$ plane at 99 \% confidence
    level. The degree of fine-tuning to the quadratic coupling of the
    Higgs field is also shown as light shaded regions. Cross marks 1,2, and 3
    are representative points for our simulation study.}
  \label{fig:constraints}
 \end{center}
\end{figure}

The result is shown in Fig.\ \ref{fig:constraints}, where the
constraints on $f$ and $\lambda_2$ at 99\% confidence level ($\chi^2
=$ 11.34) are depicted. The region $\lambda_2 < 1$ is not favored due
to electroweak precision measurements, because a large mixing angle
between $t$ and $T_+$ is predicted in this region, which leads to a
significant contribution to the custodial symmetry breaking. The
region $f < 570$ GeV, which corresponds to $m_{A_H} < m_W$, is not
attractive because the pair annihilation of $A_H$ into gauge-boson
pair is kinematically forbidden. Here, we should comment on other
parameters integrated out from the probability function. It can be
easily seen that $\alpha^{-1}(m_Z)$, $G_F$, $m_Z$, and $m_t$ are
almost fixed due to the precise measurements of these
observables. Furthermore, once ($f$, $\lambda_2$) is fixed, $m_h$ is
also fixed by the WMAP observation, because the annihilation cross
section of dark matter is sensitive to $m_h$. Here and hereafter, at
each $(f,\lambda_2)$ point, we use values of these parameters which
minimize the $\chi^2$-function. The degree of fine-tuning to set the
Higgs mass on the electroweak scale is also shown in the figure. As
mentioned in the previous subsections, the quadratic coupling of the
Higgs field $\mu^2$ is generated radiatively. One of main
contributions comes from the logarithmic divergent correction of a
top-loop diagram, which yields \cite{Dobado:2006rj}
\begin{eqnarray}
  \mu^2_t
  =
  3 \frac{m_{T_+}^2}{4\pi^2}
  \frac{\lambda_1^2\lambda_2^2}{\lambda_1^2 + \lambda_2^2}
  \log\left(1 + \frac{\Lambda^2}{m_{T_+}^2}\right),
\end{eqnarray}
where $\Lambda \simeq 4\pi f$ is the cutoff scale of the model. We
used the ratio $F = 100\times(2 m_h^2)/(\mu^2_t)$ \% to estimate the
degree of fine-tuning. It can be seen that too large $f$ and
$\lambda_2$ are not attractive from the view point of the fine-tuning.

\begin{table}[t]
  \center{
    \begin{tabular}{lccc}
      \hline
      \hline
      & Point 1 & Point 2 & Point 3
      \\ \hline 
      $f$ (GeV) & 570 & 600 & 570 
      \\
      $\lambda_2$ & 1.0 & 1.1 & 1.4 
      \\
      $\sin\beta$ & 0.20 & 0.16 & 0.11
      \\
      $m_h$ (GeV) & 145 & 131 & 145 
      \\
      $m_{A_H}$ (GeV) & 80.1 & 85.4 & 80.1 
      \\
      $m_{T_-}$ (GeV) & 570 & 660 & 798 
      \\
      $m_{T_+}$ (GeV) & 772 & 840 & 914 
      \\
      $\sigma (pp \rightarrow T_- \bar{T}_- + X)$ (pb)
      & 1.26 & 0.54 & 0.17 
      \\
      $\sigma (pp \rightarrow T_+ \bar{T}_+ + X)$ (pb)
      & 0.21 & 0.13 & 0.07 
      \\
      $\sigma (pp \rightarrow T_+ + X)$ (pb)
      & 0.29 & 0.15 & 0.05
      \\
      $\sigma (pp \rightarrow \bar{T}_+ + X)$ (pb)
      & 0.14 & 0.07 & 0.02 
      \\
      ${\rm Br}(T_+ \rightarrow W^+ b)$ & 50.8 \% & 50.8 \% & 53.3 \% 
      \\
      ${\rm Br}(T_+ \rightarrow Z t)$ & 21.1 \% & 21.8 \% & 23.6 \% 
      \\
      ${\rm Br}(T_+ \rightarrow h t)$ & 15.8 \% & 17.4 \% & 19.1 \% 
      \\
      ${\rm Br}(T_+ \rightarrow T_- A_H)$ & 12.3 \% & 10.0 \% & 4.03 \% 
      \\
      \hline
      \hline
    \end{tabular}
  }
  \caption{\small Representative points used in our simulation study.}
  \label{table:points}
\end{table}

Representative points used in our simulation study are shown in Fig.\
\ref{fig:constraints} and their details can be found in Table
\ref{table:points}. Masses of $A_H$ and $T_\pm$, cross sections for
$T_{\pm}$ pair and single $T_+$ productions, and branching ratios of
$T_+$ decay are also shown in each representative point. Note that the
$T_-$ quark decays into the stable $A_H$ and the top quark with almost
100\% branching ratio.

\section{Signals from the LHT Events}
\label{sec:tplus}
\setcounter{equation}{0}

Now, we consider the $T_+$ and $T_-$ production processes and their
signals at the LHC. At the LHC, there are two types of $T_+$
production processes, pair production and single production processes,
both of which are important. Thus, in the following, we discuss these
processes separately. In addition, we also discuss the $T_-\bar{T}_-$
pair production.

\subsection{$T_+\bar{T}_+$ pair production}

First, we discuss the $T_+\bar{T}_+$ pair production process. 
Once produced, $T_+$ decays as $T_+\rightarrow bW^+$, $tZ$, $hZ$, and
$A_HT_-$. Branching ratios for individual decay modes depend on the
underlying parameters. However, in most of the cases,
$Br(T_+\rightarrow bW^+)$ becomes larger than 0.5, and many of $T_+$
decay into $bW^+$. Thus, in the experimental situation, the analysis
using the decay mode $T_+\rightarrow bW^+$ is statistically
preferred. In such a case, the $t$ quark production events
become irreducible background. We will propose a set of kinematical
cuts suitable for the elimination of background.

For the $T_+\bar{T}_+$ production process, the most dangerous
background is the $t\bar{t}$ production which has larger cross section
than the $T_+\bar{T}_+$ production\footnote{We use the leading
order calculation of the $t\bar{t}$ production cross section 
which is $460$ pb.
}. 
Thus, we need to develop
kinematical cuts to suppress the $t\bar{t}$ background.  We propose to
use the fact that the jets in the signal events are likely to be very
energetic because they are from the decay of heavy particles (i.e.,
$T_+$ or $\bar{T}_+$). Consequently, the signal events are expected
to have large $M_{\rm eff}$, which is defined by the sum of transverse
momenta of high $p_T$ objects and missing transverse momentum
$p_T^{\rm (miss)}$:
\begin{eqnarray}
  M_{\rm eff} \equiv \sum_{\rm jets} p_T
  + \sum_{\rm leptons} p_T
  + \sum_{\rm photons} p_T
  + p_T^{\rm (miss)}.
\end{eqnarray}
In our study, only the jets with $p_T>30\ {\rm GeV}$ are included into
the high $p_T$ objects in order to reduce the contamination of QCD
activities. We expect that the number of background events can be
significantly reduced once we require that $M_{\rm eff}$ be large
enough; in the following, we will see that this is indeed the case.

Once the backgrounds are reduced, the $T_+\bar{T}_+$ production events
are reconstructed relatively easily. Here, we concentrate on the
dominant decay mode $T_+\rightarrow bW^+$. Then, the signal events are
primarily from the process $pp\rightarrow T_+\bar{T}_+$, followed by
$T_+\rightarrow bW^+$ and $\bar{T}_+\rightarrow \bar{b}W^-$. In
particular, in order to constrain the mass of $T_+$, we use the
process in which one of the $W$-boson decays hadronically while the
other decays leptonically. At the parton level, the final state
consists of two $b$-jets, two quark jets from $W^\pm$, one charged
lepton and one neutrino from $W^\mp$. Thus, the signal events are
characterized by
\begin{itemize}
 \item Several energetic jets,
 \item One isolated lepton,
 \item Missing $p_T$ (due to the neutrino emission).
\end{itemize}

Using the fact that, in the signal events, the missing momentum is due
to the neutrino emission, we reconstruct two $T_+$ systems, which we
call $T_+^{\rm (lep)}$-system and $T_+^{\rm (had)}$-system; here, the
$T_+^{\rm (lep)}$-system ($T_+^{\rm (had)}$-system) consists of high
$p_T$ objects which are expected to be from $T_+$ or $\bar{T}_+$ whose
decay is followed by the leptonic (hadronic) decay of the
$W$-boson. To determine $T_+^{\rm (lep)}$- and $T_+^{\rm
  (had)}$-systems, we first assume that all the missing $p_T$ is
carried away by the neutrino. With this assumption, the neutrino
momentum $p_\nu$ (in particular, the $z$-component of $p_\nu$) is
calculated, requiring $(p_l+p_\nu)^2=m_W^2$. Then, we define $T_+^{\rm
  (lep)}$-system as the charged lepton, reconstructed neutrino, and
one of the three leading jets, while $T_+^{\rm (had)}$-system is the
rest of the high $p_T$ objects. Since there is a two-fold ambiguity in
reconstructing the neutrino momentum, there exist six possibilities in
classifying high-$p_T$ objects into $T_+^{\rm (lep)}$- and $T_+^{\rm
  (had)}$-systems. Using the fact that $T_+^{\rm (lep)}$- and
$T_+^{\rm (had)}$-systems have the same invariant mass in the ideal
case, we choose one of the six combinations with which $|M_{T_+^{\rm
    (lep)}}-M_{T_+^{\rm (had)}}|$ is minimized, where $M_{T_+^{\rm
    (lep)}}$ and $M_{T_+^{\rm (had)}}$ are invariant masses of
$T_+^{\rm (lep)}$- and $T_+^{\rm (had)}$-systems, respectively. The
distributions of the invariant masses of $T_+^{\rm (lep)}$- and
$T_+^{\rm (had)}$-systems are expected to provide information about
the $T_+$ mass.

\begin{figure}
  \begin{center}
    \centerline{{\vbox{\epsfxsize=0.6\textwidth\epsfbox{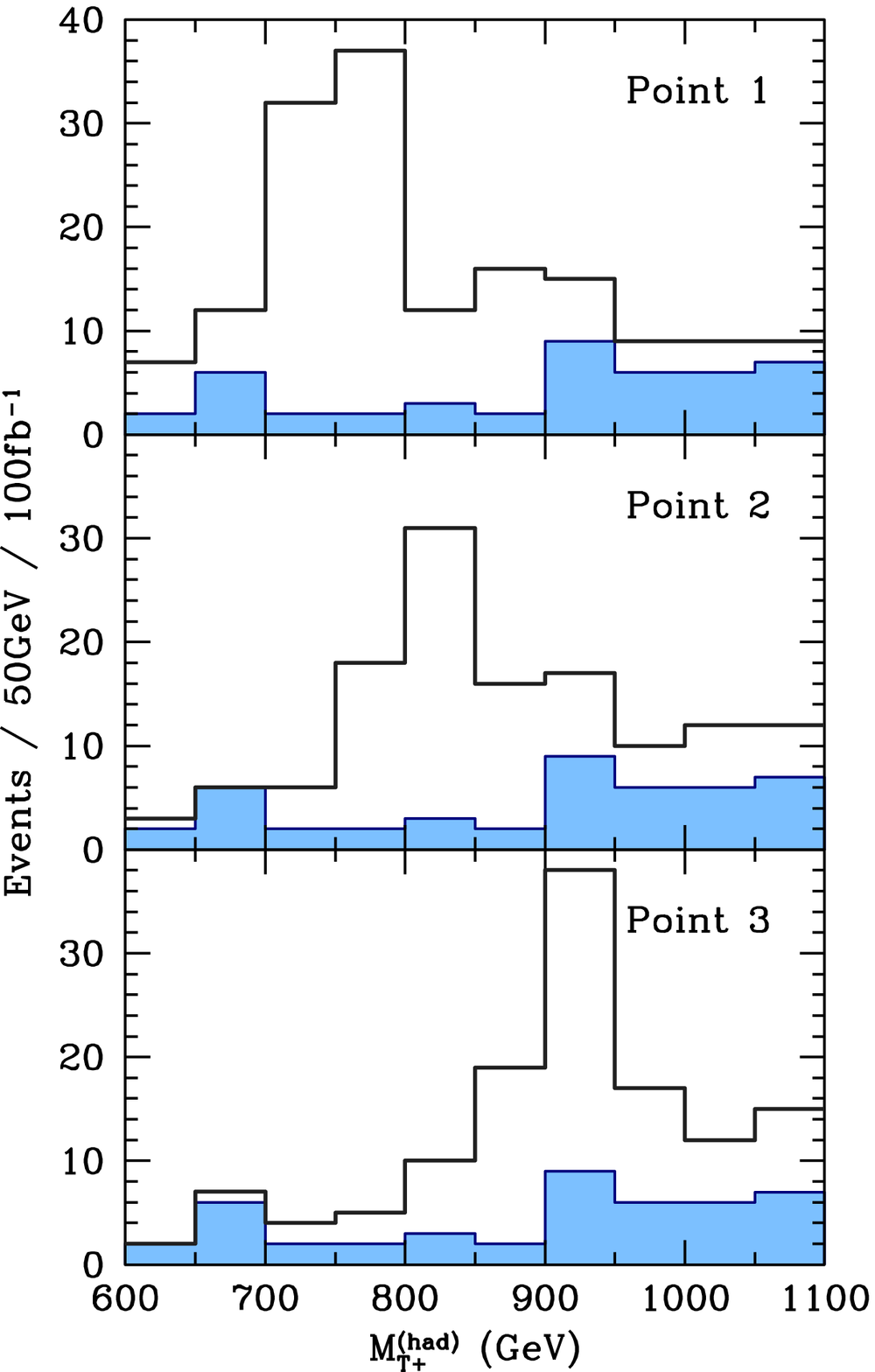}}}}
    \caption{\small Distribution of $M_{T_+^{\rm (had)}}$ for the
      Points 1 $-$ 3 (from the top to the bottom) with ${\cal L}=100\
      {\rm fb}^{-1}$. The shaded histograms are the background
      distribution, while the solid ones are for signal $+$
      background.}
    \label{fig:hist_tpl3}
    \end{center}
\end{figure}

In order to demonstrate how well our procedure works, we generate the
events for the processes $pp\rightarrow T_+\bar{T}_+$ and
$pp\rightarrow t\bar{t}$ (as well as those for $pp\rightarrow jT_+$
and $pp\rightarrow j\bar{T}_+$) with ${\cal L}=100\ {\rm
  fb}^{-1}$. The parton-level events are generated by using the
MadGraph/MadEvent packages \cite{ref:MGME}, which utilizes the HELAS
package \cite{Murayama:1992gi}. Then, Pythia package \cite{Pythia}
is used for the hadronization processes and the detector effects are
studied by using the PGS4 package \cite{PGS4}. In order to study the
$T_+\bar{T}_+$ pair production process followed by the decay processes
mentioned above, we require that the events should satisfy the
following properties:
\begin{itemize}
\item[I-0:] Three or more jets with $p_T>30\ {\rm GeV}$, and only one
  isolated charged lepton.
\end{itemize}
In addition, we adopt the following kinematical cuts:
\begin{itemize}
\item[I-1:] $p_{T,l}>50\ {\rm GeV}$ (with $p_{T,l}$ being the
  transverse momentum of the charged lepton),
 \item[I-2:] $M_{\rm eff}>1800\ {\rm GeV}$,
 \item[I-3:] $|M_{T_+^{\rm (lep)}}-M_{T_+^{\rm (had)}}|<100\ {\rm GeV}$.
\end{itemize}
Notice that the third cut is to eliminate combinatorial
backgrounds. We found that, after imposing these kinematical cuts,
events from the $jT_+$ and $j\bar{T}_+$ production processes are
completely eliminated. Then, we calculate the distributions of
$M_{T_+^{\rm (had)}}$. The results are shown in Fig.\
\ref{fig:hist_tpl3}. As one can see, the distributions have 
distinguishable peaks at around $M_{T_+^{\rm (had)}}\sim m_{T_+}$. In
addition, $t\bar{t}$ backgrounds are well below the $T_+\bar{T}_+$
signal. Thus, from the distribution of $M_{T_+^{\rm (had)}}$, we will
be able to study the properties of $T_+$.

One important observable from the distribution of $M_{T_+^{\rm
    (had)}}$ is the mass of $T_+$; once we obtain the peak of the
distribution, it will provide us an important information about
$m_{T_+}$. To see the accuracy of the determination of $m_{T_+}$, we
consider the bin $\bar{M}_{\rm bin}-\frac{1}{2}\Delta M_{\rm bin}\leq
M_{T_+^{\rm (had)}}<\bar{M}_{\rm bin}+\frac{1}{2}\Delta M_{\rm
  bin}$. Then, we calculate the number of events in the bin as a
function of the center value $\bar{M}_{\rm bin}$ with the width
$\Delta M_{\rm bin}$ being fixed. The peak of the distribution is
determined by $\bar{M}_{\rm bin}$ which maximizes the number of events
in the bin. We applied this procedure for $\Delta M_{\rm bin}=20-60\
{\rm GeV}$ (with ${\cal L}=100\ {\rm fb}$). Results for a set of
signal and background events are shown in Table\
\ref{table:peakMT+}. With repeating the Monte-Carlo (MC) analysis with
independent sets of signal samples, we found that the difference
between the position of the peak and the input value of $m_{T_+}$ is
typically $10-20\ {\rm GeV}$ or smaller. Thus, we expect a relatively
accurate measurement of $m_{T_+}$. In discussing the test of the LHT
model at the LHC, we quote $10$ and $20\ {\rm GeV}$ as the uncertainty
of $m_{T_+}$ and discuss the implication of the measurement of
$m_{T_+}$.

\begin{table}[t]
  \begin{center}
    \begin{tabular}{lccc}
      \hline\hline
      {} & {Point 1} & {Point 2} & {Point 3}\\
      \hline
      $\Delta M_{\rm bin}=30\ {\rm GeV}$ 
      & $755\ {\rm GeV}$ & $834\ {\rm GeV}$ & $913\ {\rm GeV}$ \\
      $\Delta M_{\rm bin}=40\ {\rm GeV}$ 
      & $757\ {\rm GeV}$ & $839\ {\rm GeV}$ & $918\ {\rm GeV}$ \\
      $\Delta M_{\rm bin}=50\ {\rm GeV}$ 
      & $741\ {\rm GeV}$ & $837\ {\rm GeV}$ & $910\ {\rm GeV}$ \\
      $\Delta M_{\rm bin}=60\ {\rm GeV}$ 
      & $745\ {\rm GeV}$ & $847\ {\rm GeV}$ & $912\ {\rm GeV}$ \\
      \hline\hline
    \end{tabular}
    \caption{\small Peak of the $M_{T_+^{\rm (had)}}$ distribution for
      $\Delta M_{\rm bin}=30$, $40$, $50$, and $60\ {\rm GeV}$.}
    \label{table:peakMT+}
  \end{center}
\end{table}

\subsection{Single production of $T_+$}

As well as the pair production, the single production processes
$pp\rightarrow jT_+$ and $j\bar{T}_+$ have sizable cross sections at
the LHC. (Here, $j$ denotes light quark jets.) Such processes were
discussed in \cite{Azuelos:2004dm} in the framework of the original
littlest Higgs model without the T-parity, which pointed out that the
discovery of $T_+$ may be possible by using this process. (See also
\cite{Cao:2006wk}.) Here, we reconsider the single production process for
the test of the LHT model.

So far, we have discussed that the information about the mass of $T_+$
can be obtained by studying the $T_+\bar{T}_+$ pair
production. Concerning the property of $T_+$, another important
parameter is the mixing angle $\beta$, which determines the
interaction between $T_+$ and weak bosons (i.e., $W^\pm$ and $Z$).
Importantly, the cross sections for the processes $pp\rightarrow jT_+$
and $j\bar{T}_+$ are strongly dependent on $\beta$. In particular,
since these processes are dominated by the $t$-channel $W^\pm$-boson
exchange diagram (with the use of $b$- or $\bar{b}$-quark in the
initial-state protons), the cross sections are approximately
proportional to $\sin^2\beta$. Thus, if the cross sections of the
single production processes are measured, it provides an information
about the mixing angle $\beta$. Although $pp\rightarrow jT_+$ and
$j\bar{T}_+$ have different cross section, their event shapes are very
similar (if we neglect the charges of high $p_T$ objects).  In the
following, we consider how we can measure the total cross section
$\sigma_{pp\rightarrow jT_+}+\sigma_{pp\rightarrow j\bar{T}_+}$.

As we have already discussed, once produced, $T_+$ dominantly decays
into $b$ and $W^+$. Thus, if we consider the leptonic decay of $W^+$,
there exist two energetic quarks and one charged lepton (as well as
neutrino) at the parton level in the final state. Since the mass of
$T_+$ is relatively large, the $b$-jet is expected to be very
energetic in this case. Thus, if we limit ourselves to the cases with
the leptonic decay of $W^+$, the single production events are
characterized by
\begin{itemize}
\item Two (or more) jets, one of which is very energetic (due to the
  $b$-jet),
\item One isolated lepton,
\item Missing $p_T$ (due to the neutrino emission).
\end{itemize}
As we will see, the cross section of the background events are
relatively large, so it is necessary to find a useful cut to eliminate
the backgrounds as much as possible.

One of the possible cuts is to use the invariant mass of the
``$bW^\pm$'' system. In the signal event, the dominant source of the
missing transverse momentum is the neutrino emission by the decay of
$W^+$. Thus, as we have discussed in the study of the $T_+\bar{T}_+$
pair production process, we can reconstruct the momentum of neutrino
(and hence that of $W^+$). Then, we can calculate the invariant mass
of the $bW^\pm$ system. In such a study, we presume that the highest
$p_T$ jet is the $b$-jet because, at least at the parton level, the
transverse momentum of the $b$-quark from the decay of $T_+$ is much
larger than that of the extra quark. Then, since we expect that the
mass of $T_+$ is well understood by the study of $T_+\bar{T}_+$ pair
production process, as discussed in the previous subsection, we only
use the events with relevant value of the invariant mass to improve
the signal-to-background ratio.

To estimate how well we can determine the cross section of the single
production process, we generate the signal and background events for
${\cal L}=100\ {\rm fb}^{-1}$. In \cite{Azuelos:2004dm}, it was
pointed out that the most serious backgrounds are from $t\bar{t}$
production process as well as from the single production of the
top-quark. Thus, in our study, we take account of these backgrounds.

Once the event samples are generated, we require the following event
shape:
\begin{itemize}
\item[II-0:] The number of isolated lepton is 1, the number of jets
  (with $p_T>30\ {\rm GeV}$) is 2.
\end{itemize}

In the next step, as in the case of the $T_+\bar{T}_+$ pair
production, we reconstruct the momentum of the neutrino assuming that
the transverse momentum of the neutrino is given by the observed
missing $p_T$. In reconstructing the neutrino momentum $p_\nu$, there
exists two-fold ambiguity; we denote the reconstructed neutrino
momenta $p_\nu^{(i)}$ ($i=1,2$). For each reconstructed momentum, we
calculate the invariant mass of the $bW$ system:
\begin{eqnarray}
  M_{bW}^{(i)} = \sqrt{ \left(p_{j1} + p_l + p_\nu^{(i)} \right)^2 },
\end{eqnarray}
postulating that the highest $p_T$ jet corresponds to the $b$
jet. Even though one of $M_{bW}^{(i)}$ is with the wrong
$p_\nu^{(i)}$, we found that, in the signal event, the typical
difference between $M_{bW}^{(1)}$ and $M_{bW}^{(2)}$ are relatively
small compared to that in the background events. Thus, we reject the
events unless $|M_{bW}^{(1)}-M_{bW}^{(2)}|$ is small enough.

We also comment on another useful cut to eliminate the $t\bar{t}$
background. In the $t\bar{t}$ background events, the highest $p_T$ jet
is likely to be from the overlapping of several hadronic objects from
different partons if the $p_T$ is required to be very large. In our
analysis, the cone algorithm (with $\Delta R=0.5$) is used to identify
isolated jets. Then, if several partons from the decay of top quark or
$W$-boson are emitted in almost the same direction, hadronized objects
from those partons are grouped into a single jet, which may be
identified as the $b$-originated jet in the present analysis. One of the
method to reject such a background is to use the jet-mass variable,
which is the invariant mass of the jet constructed from all the
(observed) energy and momentum that are contained in the jet. The
jet-mass of such a jet is likely to be much larger than that of the
$b$-jet. As we will show, the number of background from the $t\bar{t}$
production process is significantly reduced if the jet mass is
required to be small enough.

Now, we show the results of our MC analysis. In our analysis, we use
the following kinematical cuts:
\begin{itemize}
\item[II-1:] $p_{T,l}>100\ {\rm GeV}$, $p_T^{\rm (miss)}>100\ {\rm
    GeV}$,
\item[II-2:] $p_{T,j1}>300\ {\rm GeV}$, and $M_{j1+j2}>500\ {\rm
    GeV}$, with $M_{j1+j2}$ being the invariant mass of total jets,
\item[II-3:] $M_{j1}<50\ {\rm GeV}$, with $M_{j1}$ being the jet mass
  of the leading jet,
\item[II-4:] $|M_{bW}^{(1)}-M_{bW}^{(2)}|<50\ {\rm GeV}$.
\end{itemize}

\begin{figure}
  \begin{center}
    \centerline{{\vbox{\epsfxsize=0.6\textwidth\epsfbox{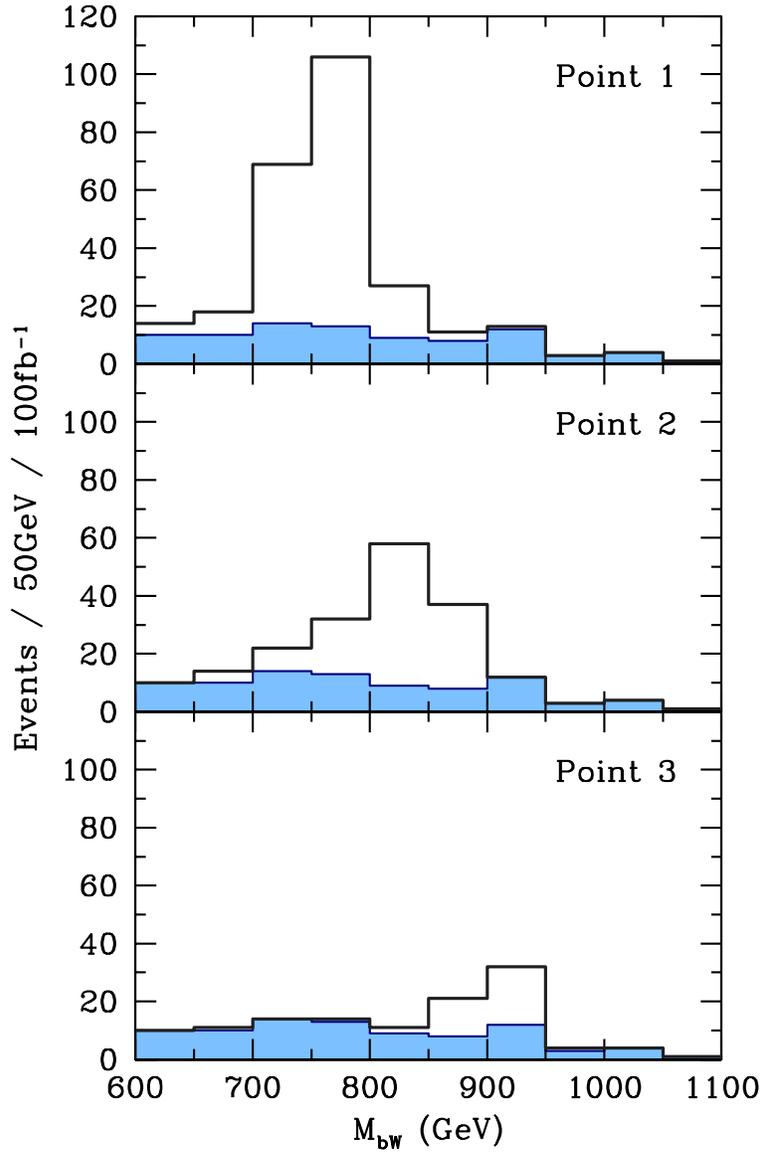}}}}
    \caption{\small Distribution of $M_{bW}$ for the Points 1 $-$ 3
      (from the top to the bottom). The shaded histograms are the
      background distribution, while the solid ones are for signal $+$
      background.}
    \label{fig:hist_jtpl3}
  \end{center}
\end{figure}

In Fig.\ \ref{fig:hist_jtpl3}, we plot the distribution of the
``averaged'' invariant mass of the $bW$ system:
\begin{eqnarray}
 M_{bW} \equiv \frac{1}{2} \left( M_{bW}^{(1)} + M_{bW}^{(2)} \right).
\end{eqnarray}
As one can see, the distribution from the signal events is peaked at
around $M_{bW}\sim m_{T_+}$, while the background distribution is
rather flat. In addition, at around $M_{bW}\sim m_{T_+}$, the number
of signal events becomes significantly larger than that of background
in particular when the parameter $\sin\beta$ is relatively large. In
such a case, the number of the single production events can be
extracted from the distribution by using, for example, the side-band
method\footnote{It should be also possible to constrain the mass of
  $T_+$ from the peak of the distribution of $M_{bW}$. In this paper,
  we will not discuss such a possibility.}.

In Table\ \ref{table:number_singleT+}, with the data for the Point 2,
we show the number of events in the event region, which we define
$m_{T_+}-50\ {\rm GeV}\leq M_{bW}\leq m_{T_+}+50\ {\rm GeV}$, and
those in the sidebands, $m_{T_+}-150\ {\rm GeV}\leq M_{bW}\leq
m_{T_+}-50\ {\rm GeV}$ and $m_{T_+}+50\ {\rm GeV}\leq M_{bW}\leq
m_{T_+}+150\ {\rm GeV}$, after imposing the kinematical cuts mentioned
above. Assuming that the numbers of signal and background events in
the signal region are determined by using the sideband events, and
that the cross section for the single $T_+$ production process can be
obtained from the number of events in the signal region, the single
$T_+$ production cross section may be determined with the uncertainty
of $10-20\ \%$. (The uncertainty here is statistical only.)

Using the result of the $m_{T_+}$ determination with the
$T_+\bar{T}_+$ pair production process, the information about the
cross section can be converted to that of the mixing angle $\beta$. If
the uncertainties in the theoretical calculation of the cross sections
are under control, we obtain a constraint on $\beta$. Since the cross
section for the single production process is proportional to
$\sin^2\beta$, $\sin\beta$ is determined with the
accuracy of $5-10\ \%$ if the cross section is determined with the
accuracy of $10-20\ \%$\footnote{The cross section also depend on the
  mass of $T_+$. Thus, the constraint on the cross section should
  provide a constraint on the $\beta$ vs.\ $m_{T_+}$ plane. In our
  discussion, for simplicity, we only consider the $\beta$ dependence
  of the cross section by using the fact that the mass of $T_+$ is
  expected to be determined from the $T_+\bar{T}_+$ pair production
  process.}. In the next section, we discuss the implication of the
determination of $\beta$ at this level in testing the LHT model.

\begin{table}[t]
  \begin{center}
    \begin{tabular}{lrrrrrrrrr}
      \hline\hline
      {}
      & \multicolumn{3}{c}{Lower Sideband} 
      & \multicolumn{3}{c}{Event Region} 
      & \multicolumn{3}{c}{Upper Sideband}\\
      & Signal & $t\bar{t}$ & $jt + j\bar{t}$
      & Signal & $t\bar{t}$ & $jt + j\bar{t}$
      & Signal & $t\bar{t}$ & $jt + j\bar{t}$\\
      \hline
      II-0
      & 313 & 21706 & 13509 & 522 & 12585 & 8609 & 116 & 7810 & 5362 \\
      II-0, 1
      & 108 &  3366 &   376 & 234 &  2352 &  363 &  44 & 1747 &  237 \\
      II-0, 1, 2
      &  45 &   428 &    53 & 144 &   446 &   76 &  14 &  440 &   86 \\
      II-0, 1, 2, 3
      &  30 &    30 &    47 & 114 &    27 &   50 &   8 &   21 &   69 \\
      II-0, 1, 2, 3, 4
      &  21 &    12 &    18 &  84 &    11 &   12 &   2 &    3 &   16 \\
      \hline\hline
    \end{tabular}
    \caption{\small Number of the signal events/$t\bar{t}$
      background/single top background events in the event region
      ($m_{T_+}-50\ {\rm GeV}\leq M_{bW}\leq m_{T_+}+50\ {\rm GeV}$)
      as well as in the lower and upper sidebands ($m_{T_+}-150\ {\rm
        GeV}\leq M_{bW}\leq m_{T_+}-50\ {\rm GeV}$ and $m_{T_+}+50\
      {\rm GeV}\leq M_{bW}\leq m_{T_+}+150\ {\rm GeV}$,
      respectively). Point 2, where $m_{T+}=840\ {\rm GeV}$, is used.}
    \label{table:number_singleT+}
  \end{center}
\end{table}

Before closing this subsection, we comment on the uncertainties which
we have neglected so far. As we have mentioned, the single production
process occurs by using the $b$ or $\bar{b}$ in the sea quark of the
initial-state proton. Thus, for the theoretical calculation of the
cross sections, it is necessary to understand the parton distribution
functions for the $b$ and $\bar{b}$ quarks (as well as those of
lighter quarks). Information about the parton distributions of the
$b$-quark may be obtained by using the single top (and anti-top)
productions. As we have seen, significant amount of single top
productions occur at the LHC (which has been seen to be one of the
dominant backgrounds to the single $T_+$ production process). Since
the single top production also occurs by using the $b$ quark in
proton, information about the parton distribution function of $b$ will
be obtained by studying the single top production process. In this
paper, we do not go into the detail of such study, but we just assume
that the parton distribution function of $b$ will become available
with small uncertainty once the LHC experiment will start. We also
note here that it is also important to understand the efficiency to
accept the single production events (as well as the background events)
after the cuts, whose uncertainties have been neglected in our
discussion.

\subsection{$T_-\bar{T}_-$ pair production}

For the study of the LHT model at the LHC, it is also relevant to
consider the $T$-odd top partner, $T_-$, and the lightest $T$-odd
particle, $A_H$. For the study of $T$-odd particles, it is important
to consider the $T_-\bar{T}_-$ pair production process, which was
discussed in \cite{Matsumoto:2006ws, Nojiri:2008ir}. Here, we
reconsider the importance of this process for the test of the LHT
model.

At the LHC, $T_-$ is pair produced via $pp\rightarrow T_-\bar{T}_-$,
then decays as $T_-\rightarrow tA_H$. Since $A_H$ is undetectable, the
$T_-$ production events always result in missing $p_T$ events and
hence the direct measurements of the masses of $T_-$ and $A_H$ are
difficult.

One powerful method to study $m_{T_-}$ and $m_{A_H}$ is the so-called
$M_{T2}$ analysis \cite{MT2}, combined with the hemisphere analysis
\cite{hemisphere}. If the $t$ and $\bar{t}$ systems are somehow
reconstructed, one can constrain $m_{T_-}$ and $m_{A_H}$ from the
distribution of the so-called $M_{T2}$ variable. For the event
$pp\rightarrow T_-\bar{T}_-$ followed by $T_-\rightarrow tA_H$ and
$\bar{T}_-\rightarrow \bar{t}A_H$, the $M_{T2}$ variable is defined as
\begin{eqnarray}
  M_{T2}^2 (\tilde{m}_{A_H}) = 
  \min_{{\bf p}^t_{\rm T} + {\bf q}^{\bar{t}}_{\rm T} 
    + {\bf p}^{A_H}_{\rm T} + {\bf q}^{A_H}_{\rm T}=0}
  \left[
    \max \left\{
    M_{\rm T}^2 ({\bf p}^t_{\rm T}, {\bf p}^{A_H}_{\rm T};
    \tilde{m}_{A_H}),
    M_{\rm T}^2 ({\bf q}^{\bar{t}}_{\rm T}, {\bf q}^{A_H}_{\rm T};
    \tilde{m}_{A_H})
    \right\} \right],
  \label{MT2^2}
\end{eqnarray}
where the transverse mass $M_{\rm T}$ is defined as
\begin{eqnarray}
  M_{\rm T} ({\bf p}^t_{\rm T}, {\bf p}^{A_H}_{\rm T}; 
  \tilde{m}_{A_H})
  =
  \sqrt{ ( | {\bf p}^t_{\rm T} |^2 + m_t^2 )
    ( | {\bf p}^{A_H}_{\rm T} |^2
    + \tilde{m}_{A_H}^2 ) 
    - {\bf p}^t_{\rm T} {\bf p}^{A_H}_{\rm T} },
\end{eqnarray}
with $\tilde{m}_{A_H}$ being the postulated mass of $A_H$ to calculate
$M_{T2}$. In the above expression, ${\bf p}^t_{\rm T}=(p^t_x,p^t_y,0)$
and ${\bf q}^{\bar{t}}_{\rm T}=(q^{\bar{t}}_x,q^{\bar{t}}_y,0)$ are
transverse momenta of $t$ and $\bar{t}$, respectively, which are
obtained from the reconstructed top systems. The reconstruction of the
top systems is possible with sizable efficiency by using the
hemisphere method \cite{Nojiri:2008ir}. In addition, ${\bf
  p}^{A_H}_{\rm T}$ and ${\bf q}^{A_H}_{\rm T}$ are postulated
transverse momenta of the final-state $A_H$ particles, which satisfy
\begin{eqnarray}
  {\bf p}^t_{\rm T} + {\bf q}^{\bar{t}}_{\rm T} 
  + {\bf p}^{A_H}_{\rm T} + {\bf q}^{A_H}_{\rm T} = 0.
\end{eqnarray}
In the calculation of $M_{T2}$, ${\bf p}^{A_H}_{\rm T}$ and ${\bf
  q}^{A_H}_{\rm T}$ are varied under the above constraint to minimize
the quantity in the square bracket of Eq.\ (\ref{MT2^2}).

The important property of the $M_{T2}$ variable is that, if
$\tilde{m}_{A_H}$ is equal to $m_{A_H}$, the upper end-point of the
distribution of $M_{T2}$ is given by $m_{T_-}$\footnote{For a general
  value of $\tilde{m}_{A_H}$, the upper end-point of the $M_{T2}$
  distribution is given by
  \begin{eqnarray*}
    M_{T2}^{\rm (max)} (\tilde{m}_{A_H}) = 
    \frac{m_{T_-}^2 + m_t^2 - m_{A_H}^2}{2 m_{T_-}}
    + \sqrt{ 
      \left( \frac{m_{T_-}^2 + m_t^2 - m_{A_H}^2}{2 m_{T_-}} \right)^2
      + \tilde{m}_{A_H}^2 - m_t^2}.
  \end{eqnarray*}
  This can be used to check the validity of the MC analysis.}.  Thus,
once many samples of $T_-\bar{T}_-$ production events become available
at the LHC, it will be possible to determine the distribution of the
$M_{\rm T2}$ variable for each value of $\tilde{m}_{A_H}$.
The distribution of the $M_{T2}$ variable for the $T_-\bar{T}_-$
production process was studied in \cite{Nojiri:2008ir} with the choice
of $\tilde{m}_{A_H}=m_{A_H}$. In our discussion, we use the $M_{T2}$
analysis to constrain $m_{A_H}$ and $m_{T_-}$, so it is necessary to
study the distribution of the $M_{T2}$ variable for various values of
$\tilde{m}_{A_H}$.

\begin{figure}[t]
  \begin{center}
    \centerline{{\vbox{\epsfxsize=0.6\textwidth\epsfbox{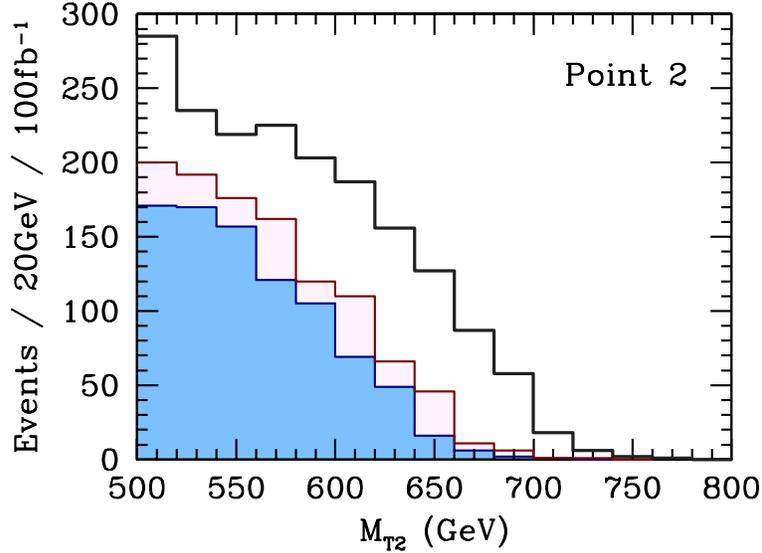}}}}
    \caption{\small Distribution of the $M_{T2}$ variable for
      $\tilde{m}_{A_H}=0$ (darkly shaded: blue), $100\ {\rm GeV}$
      (lightly shaded: pink), and $200\ {\rm GeV}$ (solid line).}
    \label{fig:hist_mt2pt2}
    \end{center}
\end{figure}

To see how the distribution of the $M_{T2}$ variable depends on
$\tilde{m}_{A_H}$, we generate the $T_-\bar{T}_-$ events (as well as
$t\bar{t}$ backgrounds) and derive the distribution of $M_{T2}$. Here,
we intend to use the events:
\begin{eqnarray}
  pp \rightarrow T_- \bar{T}_- \rightarrow t A_H \bar{t} A_H
  \rightarrow b W^+ \bar{b} W^- A_H A_H
  \rightarrow b q q'\bar{b} q'' q''' A_H A_H,
\end{eqnarray}
and we adopt the kinematical cuts used in \cite{Nojiri:2008ir}:
\begin{itemize}
 \item[III-0:] No isolated leptons,
 \item[III-1:] $p_T^{\rm (miss)}>200\ {\rm GeV}$, and $p_T^{\rm
     (miss)}> 0.2 M_{\rm eff}$.
\end{itemize}
Notice that large missing $p_T$ is expected due to the emission of two
$A_H$ particles.  Then, in order to reconstruct two top systems, we
use the hemisphere analysis with which all the high $p_T$ objects are
assigned to one of two hemispheres, $H_1$ and $H_2$, so that
\begin{eqnarray}
  \left\{ \begin{array}{ll}
  d (p_{H_1}, p_i) < d (p_{H_2}, p_i) & ~:~
  {\rm for}\ \forall i \in H_1\\
  d (p_{H_2}, p_i) < d (p_{H_1}, p_i) &~:~
  {\rm for}\ \forall i \in H_2
  \end{array} \right. ,
\end{eqnarray}
where $p_{H_I}$ is the momentum of the $I$-th hemisphere which is
defined as
\begin{eqnarray}
  p_{H_I} = \sum_{i \in H_I} p_i,
\end{eqnarray}
and
\begin{eqnarray}
  d (p_{H_I}, p_i) =
  \frac{(E_{H_I} - |{\bf p}_{H_I}| \cos\theta_{Ii}) E_{H_I}}
  {(E_{H_I} + E_i)^2},
\end{eqnarray}
with $\theta_{Ii}$ being the angle between ${\bf p}_{H_I}$ and ${\bf
  p}_i$. (For the details to construct the hemispheres, see
\cite{Matsumoto:2006ws}.) In the following, the first hemisphere is
defined as the one which contains the leading jet.  Once two
hemispheres are determined, we impose the following cuts to eliminate
backgrounds:
\begin{itemize}
\item[III-2:] Numbers of jets (with $p_T>30\ {\rm GeV}$) in $H_1$ and
  $H_2$ are either equal to or smaller than 3.
\item[III-3:] $p_{T,H_I}>200\ {\rm GeV}$ ($I=1$, $2$), where
  $p_{T,H_I}$ is the transverse momentum of the hemisphere $H_I$.
\item[III-4:] $50\ {\rm GeV}\leq M_{H_I} \leq 190\ {\rm GeV}~(I=1,2)$,
  where $M_{H_I}$ is the invariant mass of the $I$-th hemisphere
   (i.e.,  $M_{H_I}=\sqrt{p_{H_I}^2}$).
\end{itemize}
As shown in \cite{Nojiri:2008ir}, with the cuts III-0 $-$ III-3, peaks
around $\sim m_t$ are obtained in the distributions of $M_{H1}$ and
$M_{H2}$.
Then, postulating that the momenta of $t$ and $\bar{t}$ are given by
those of two hemispheres, we calculate the distribution of the
$M_{T2}$ variable defined in Eq.\ (\ref{MT2^2}) for several values of
the postulated mass $\tilde{m}_{A_H}$. Here, we use the underlying
parameters for the Point 2. The results for $\tilde{m}_{A_H}=0$, $100\
{\rm GeV}$, and $200\ {\rm GeV}$, for which the theoretically expected
end-points are $648\ {\rm GeV}$, $664\ {\rm GeV}$, and $708\ {\rm
  GeV}$, respectively, are shown in Fig.\ \ref{fig:hist_mt2pt2}. Here,
the distributions shown in the figure include contributions from the
$t\bar{t}$ background; however, we have checked that there is no
contamination of the $t\bar{t}$ events at the end-point region.

As one can see, the position of the upper end-point changes
consistently with the theoretical value of the end-point. Thus, by
using the $M_{T2}$ variable, we expect to obtain a constraint on the
$m_{A_H}$ vs.\ $m_{T_-}$ plane, which can be transferred to a
constraint on the $\lambda_2$ vs.\ $f$ plane. In order to derive the
constraint, it is necessary to understand how well the position of the
upper end-point can be determined.  Detailed properties of the
distribution of the $M_{T2}$ variable should depend on the kinematical
cuts as well as on the detector performances.  An extensive study of
the fitting function to determine the end-point is beyond the scope of
this paper. Here, we simply use the quadratic function to estimate the
end-point. For example, for $\tilde{m}_{A_H}=100\ {\rm GeV}$ (for
which the theoretical prediction of the end-point is $664\ {\rm
  GeV}$), the end-point is estimated as $M_{T2}^{\rm (max)}=(664\pm
9)\ {\rm GeV}$ ($M_{T2}^{\rm (max)}=(676\pm 3)\ {\rm GeV}$) using the
data with $550\ {\rm GeV}\leq M_{T2}\leq 650\ {\rm GeV}$ ($580\ {\rm
  GeV}\leq M_{T2}\leq 680\ {\rm GeV}$). Thus, in the following
discussion, we adopt the error of $10 - 20 {\rm GeV}$
in the determination of the end-point, although a better result may be
possible if a detailed study of the shape of the end-point is
performed.

\section{Test of the LHT Model}
\label{sec:test}
\setcounter{equation}{0}

Now we discuss how and how well we can test the LHT model using the
results obtained in the previous section. As we discussed in Section\
\ref{sec:model}, the LHT model is parametrized by two parameters, $f$
and $\lambda_2$. Thus, if there exists three or more observables, a
non-trivial test becomes possible.

In the following, we adopt the Point 2 as the underlying parameter
point, and assume that $m_{T_+}$, $\sin\beta$, and the end-point
of the $M_{T2}$ variable can be experimentally determined as
\begin{eqnarray}
  m_{T_+} &=& [ m_{T_+} ]_{\rm Point\ 2} \pm \delta m_{T_+},
  \label{const_mT+}
  \\
  \sin\beta &=& [ \sin\beta ]_{\rm Point\ 2} 
  \pm \delta \sin\beta,
  \label{const_sinbeta}
  \\
  M_{T2}^{\rm (max)} &=& [ M_{T2}^{\rm (max)} ]_{\rm Point\ 2}
  \pm \delta M_{T2}^{\rm (max)},
  \label{const_MT2}
\end{eqnarray}
where $[\cdots]_{\rm Point\ 2}$ denotes the value in Point 2.  From
the discussion in the previous section, we adopt the following
uncertainties of the quantities mentioned above:
\begin{itemize}
\item Case 1:
  \begin{eqnarray}
    \delta m_{T_+} &=& 20\ {\rm GeV},
    \\
    \delta \sin\beta/\sin\beta &=& 10\ \%,
    \\
    \delta M_{T2}^{\rm (max)} &=& 20\ {\rm GeV},
  \end{eqnarray}
\item Case 2:
  \begin{eqnarray}
    \delta m_{T_+} &=& 10\ {\rm GeV},
    \\
    \delta \sin\beta/\sin\beta &=& 5\ \%,
    \\
    \delta M_{T2}^{\rm (max)} &=& 10\ {\rm GeV},
  \end{eqnarray}
\end{itemize}
In the Case 2, smaller uncertainties are adopted compared to the Case
1.

\begin{figure}[t]
  \begin{center}
    \centerline{{\vbox{\epsfxsize=0.6\textwidth\epsfbox{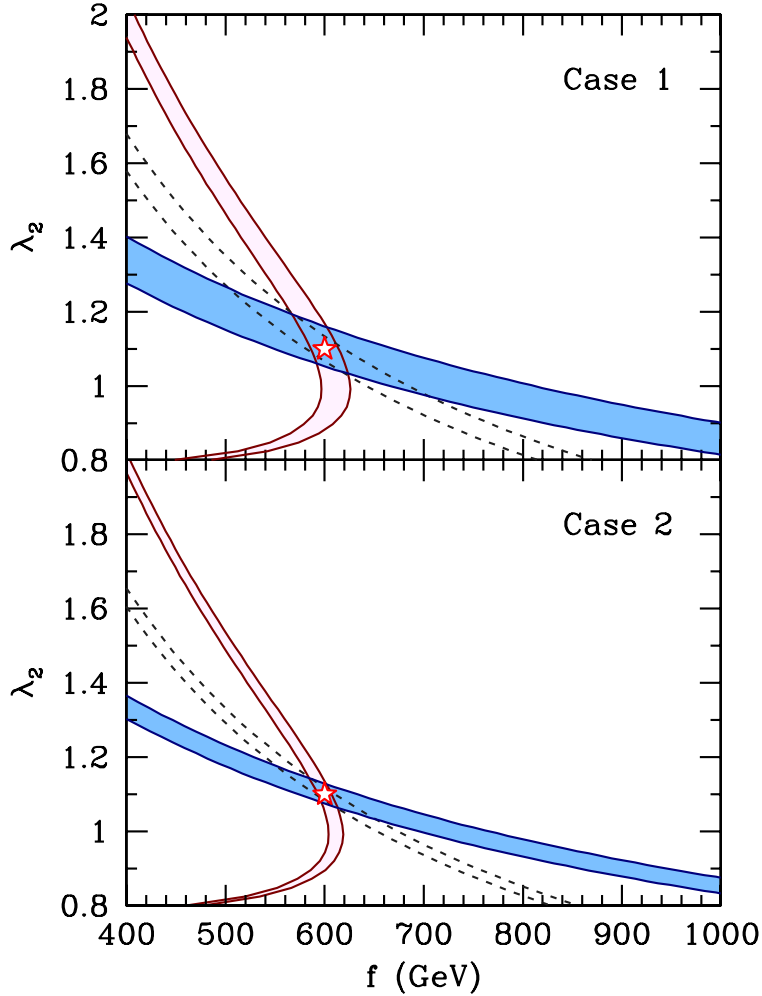}}}}
    \caption{\small Expected constraints on the $f$ vs.\ $\lambda_2$
      plane for Cases 1 and 2 (upper and lower, respectively). Point
      2 is used as the underlying parameter point. Constraints from
      the measurements of $m_{T_+}$, $\sin\beta$, and $M_{T2}^{\rm
        (max)}$ are given by the lightly-shaded (pink) region, region
      between the dashed lines, and darkly-shaded (blue) region,
      respectively. The star in the figure is the underlying point.}
    \label{fig:fvslam2}
    \end{center}
\end{figure}

In Fig.\ \ref{fig:fvslam2}, we show the allowed region on the $f$ vs.\
$\lambda_2$ plane for the Cases 1 and 2. As one can see, measurements
of $m_{T_+}$, $\sin\beta$, and $M_{T2}^{\rm (max)}$ provide three
different constraints on the $f$ vs.\ $\lambda_2$ plane. It should be
noticed that, because each of the constraints gives a narrow band on
the $f$ vs.\ $\lambda_2$ plane, we can quantitatively test if the
observed signals are consistent with the predictions of the LHT model;
if the three bands meet at a single point, as shown in Fig.\
\ref{fig:fvslam2}, it gives a quantitative confirmation of the LHT
model.

It is also notable that the measurements of $m_{T_+}$, $\sin\beta$,
and $M_{T2}^{\rm (max)}$ give accurate determinations of $f$ and \
$\lambda_2$. For example, reading the lower and upper bounds on these
parameters from the allowed region in the Case 1 (Case 2), we obtain
the constraints $566\ {\rm GeV}< f< 624\ {\rm GeV}$ and
$1.03<\lambda_2<1.20$ ($584\ {\rm GeV}< f< 613\ {\rm GeV}$ and
$1.06<\lambda_2<1.15$). One of the implications is that, with the
determination of $f$, we can also determine $m_{A_H}$ in the LHT
model. (See Eq.\ (\ref{m(A_H)}).) Since $A_H$ is a very weakly
interacting particle, the direct determination of its mass is
difficult as discussed in the previous section. Thus, the
determination of $f$ gives an important information about $m_{A_H}$.

Finally, we discuss an implication to cosmology. $A_H$ is a viable
candidate of dark matter. The thermal relic density of $A_H$ strongly
depends on the pair annihilation cross section of $A_H$; in the
present case, $A_H$ pair-annihilates into weak boson pair via the
$s$-channel exchange of the Higgs boson. The pair annihilation cross
section is obtained once $f$ and $m_h$ are known. As we have
discussed, $f$ can be determined with the studies of the top
partners. In addition, at the LHC, it is expected that the Higgs boson
will be found and its properties will be studied in detail. For
example, if $m_h=130-150\ {\rm GeV}$, the Higgs mass will be
determined with the uncertainty of $\sim 200\ {\rm MeV}$
\cite{AtlasTdr, CmsTdr}\footnote{For a discussion of Higgs
  phenomenology in the LH models, see \cite{Chen:2006cs}.}; in the
following, we assume that the Higgs mass can be determined with the
accuracy of $200\ {\rm MeV}$ at the Point 2. Then, combining the
information about the top-partners and the Higgs boson from the LHC,
it will become possible to reconstruct the thermal relic density of
$A_H$. Comparison of the theoretically calculated relic density and
observed dark matter density provides an important test of the
cosmological scenario in the framework of the LHT model; if the
theoretical prediction of the relic density is consistent with the
dark matter density observed, it will be a strong indication of the
scenario where $A_H$ is dark matter.

\begin{figure}[t]
  \begin{center}
    \centerline{\epsfxsize=0.6\textwidth\epsfbox{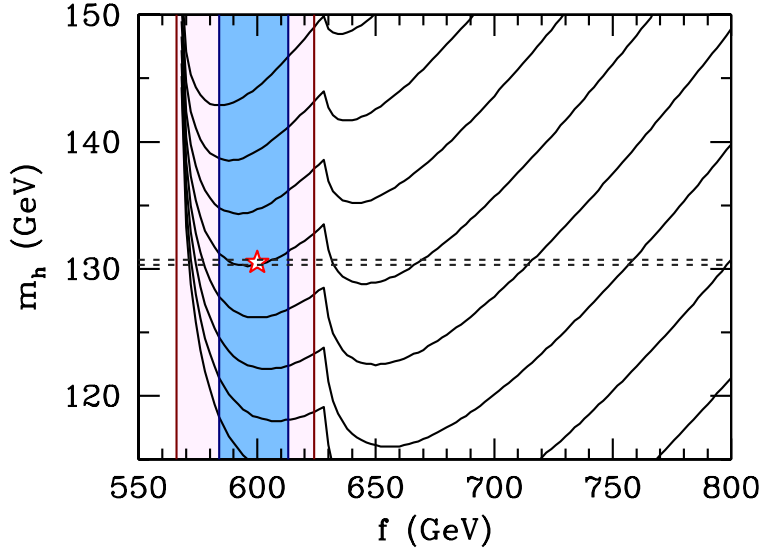}}
    \caption{\small Contours of constant $\Omega_{A_H}h^2$ on $f$
      vs. $m_h$ plane. Contours are for $\Omega_{A_H}h^2=0.06$,
      $0.08$, $0.10$, $0.12$, $0.14$, $0.16$, $0.18$, and $0.20$ from
      above. Expected bound on $f$ is shown in the shaded region;
      lightly-shaded (pink) region is for the Case 1 while the
      darkly-shaded (blue) region is for the Case 2. The dotted lines
      are the expected upper and lower bounds on the Higgs mass.}
    \label{fig:lhomega}
    \end{center}
\end{figure}

To see how well we can perform this test, we calculate the thermal
relic density $\Omega_{A_H}$; the contours of constant
$\Omega_{A_H}h^2$ (with $h$ being the Hubble constant in units of
km/sec/Mpc) are shown in Fig.\ \ref{fig:lhomega} on $f$ vs.\ $m_h$
plane. When $f\lesssim 570\ {\rm GeV}$, $A_H$ becomes lighter than
$W^\pm$. In such a case, the pair annihilation cross section of $A_H$
is extremely suppressed, resulting in very large value of
$\Omega_{A_H}h^2$. On the contrary, for $f\gtrsim 570\ {\rm GeV}$,
$\Omega_{A_H}h^2$ is found to have mild dependence on $f$ and
$m_h$. In the same figure, we also show the expected constraints on
$f$ and $m_h$. As one can see, determination of the $f$ parameter
plays an important role in reconstructing the dark matter density. In
particular, we can see that, combined with the precise measurement of
the Higgs mass, $\Omega_{A_H}h^2$ can be reconstructed very accurately
in the Case 2 where the masses of top partners and mixing parameter
$\beta$ are well determined; with the determination of $m_h$ and $f$
for the Case 2 mentioned above, the density parameter is constrained
to be $0.118< \Omega_{A_H}h^2<0.126$. (The underlying value of
$\Omega_{A_H}h^2$ is $0.120$.) On the contrary, in the Case 1 where
the uncertainty in $f$ is relatively large, bound on the density
parameter is found to be $\Omega_{A_H}h^2>0.118$. Thus, in such a
case, $\Omega_{A_H}h^2$ cannot be bounded from above. This
is mainly due to the fact that we chose the underlying value of
$m_{A_H}$ close to $m_W$; with larger value of $m_{A_H}$, a better
reconstruction of $\Omega_{A_H}h^2$ is expected even with a larger
uncertainty in $f$.

\section{Summary}
\label{sec:summary}

%Solving the (little) hierarchy problem in the SM has been one of
%driving forces to consider new physics models beyond the SM. Little
%Higgs models are one of such interesting possibilities.  In little
%Higg models, the fermionic partner of top quark cancels the main
%quadratically divergent correction (induced by top quark) to the Higgs
%boson mass parameter, and hence it is an essential part of the
%models. Therefore, unveiling properties of heavy partners of top quark
%at the LHC will be important to understand the mechanism to solve the
%hierarchy problem.

In this paper, we have studied the $T_+\bar{T}_+$ pair, single-${T_+}$
and $T_-\bar{T}_-$ pair productions at the LHC in the framework of the
littlest Higgs model with T-parity, by performing a numerical
simulation on three representative points.  For $T_+\bar{T_+}$ pair
production process, the main SM background comes from $t\bar{t}$
production.  We have developed kinematical cuts to suppress the
$t\bar{t}$ background, and found that the signal events can be well
extracted from the background.  We have shown that an accurate
determination of the mass of $T_+$ is possible.  For single-$T_+$
production, we have also proposed a set of kinematical cuts to
suppress the SM backgrounds which are from $t\bar{t}$ pair production
and single-$t$ production, and shown that the signal events can be
well reconstructed. From the measurement of the single-$T_+$
production cross section as well as the measurement of $m_{T_+}$ in
the $T_+\bar{T}_+$ pair production, we can obtain the information on
the mixing parameter ($\sin\beta$) between $T_+$ and top quark.  For
$T_-\bar{T}_-$ pair production, studying the upper end-point of the
$M_{T2}$ distribution ($M_{T2}^{(\rm max)}$), a certain relation
between $m_{A_H}$ and $m_{T_-}$ is obtained.

Since the top sector in the LHT is parametrized by two parameters,
$f$ and $\lambda_2$, each measurement of these three observables
provides a relation between $f$ and $\lambda_2$.  We have shown that
the measurements of the three observables give non-trivial
determinations of the parameters $f$ and $\lambda_2$, and hence a
quantitative test of the LHT model can be performed at the LHC.

In the LHT model, $A_H$ is a viable dark matter candidate. Since the
thermal relic density of $A_H$ strongly depends on the pair
annihilation cross section of $A_H$ into weak boson pair via
$s$-channel exchange of the Higgs boson, the masses of $A_H$ and Higgs
boson are important to calculate the thermal relic density of $A_H$.
Using the facts that $A_H$ mass can be determined by the parameter
$f$, and that not only the discovery of Higgs boson but also the
measurement of the Higgs mass are expected at the LHC, we have shown
that the relic density of $A_H$ can be calculated very accurately by
using the LHC results.  This will provide an important test of the
cosmological scenario where $A_H$ becomes dark matter.

Our studies here suggest not only that the LHC has a great potential
to discover the heavy partner of top quark which is responsible for
the cancellation of the main quadratically divergent contribution to
the Higgs mass parameter, but also that the LHC can provide 
important measurements of the observables that would lead us to a
crucial tests of the LHT model.

\setcounter{equation}{0}
\noindent\\
{\it Acknowledgments:} This work is supported in part by the
Grant-in-Aid for Scientific Research from the Ministry of Education,
Science, Sports, and Culture of Japan, No.\ 19540255 (T.M.).

\end{document}